\documentstyle[aps,prl,epsf,floats,12pt]{revtex}

\draft
\begin{document}

\title{Electron--Hole Systems in Narrow Quantum Wells:\\
       Excitonic Complexes and Photoluminescence}

\author{
   \underline{John J. Quinn}$^{1}$,
   Arkadiusz W\'ojs$^{1,2}$,
   Kyung-Soo Yi$^{1,3}$, and
   Izabela Szlufarska$^{1,2}$\\[1ex]}
\address{\small\sl
   $^1$Department of Physics, 
       University of Tennessee, Knoxville, Tennessee 37996, USA \\
   $^2$Institute of Physics, 
       Wroclaw University of Technology, Wroclaw 50-370, Poland \\
   $^3$Physics Department, 
       Pusan National University, Pusan 609-735, Korea \\[3ex]}
\address{\rm\small\parbox{6.5in}{
The energy and photoluminescence (PL) spectra of a two-dimensional 
electron gas (2DEG) interacting with a valence-band hole are studied 
in the high-magnetic-field limit as a function of the filling factor 
$\nu$ and the separation $d$ between the electron and hole layers. 
For $d$ smaller than the magnetic length $\lambda$, the hole binds one 
or more electrons to form neutral ($X^0$) or charged ($X^-$) excitons, 
and PL probes the lifetime and binding energies of these complexes 
rather than the original correlations of the 2DEG. 
The low-lying states can be understood in terms of Laughlin-type 
correlations among the constituent negatively charged Fermions 
(electrons and $X^-$'s). 
For $d$ large compared to $\lambda$, the electron--hole interaction 
is not strong enough to bind a full electron, and fractionally 
charged excitons (bound states of the hole and one or more Laughlin 
quasielectrons) $h$QE$_n$ are formed. 
The PL selection rule associated with rotational invariance 
(conservation of $L$) is only weakly violated in the interacting 
plasma, and the position and oscillator strengths of PL lines can 
be predicted and compared with numerical calculations.\\[1ex]
}}
\maketitle

\section{Introduction}

In order to obtain a better understanding of the photoluminescence (PL) 
process in fractional quantum Hall (FQH) systems, it is important to 
understand the nature of the low-lying eigenstates of the interacting 
electron--hole system. 
In this note, we study the eigenstates of a system consisting of $N$ 
electrons confined to a plane $z=0$ and interacting with one another 
and with a valence-band hole ($h$) confined to a plane $z=d$, where $d$ 
is measured in units of the magnetic length $\lambda=(\hbar c/eB)^{1/2}$. 
The cyclotron energy $\hbar\omega_c$ is assumed to be much larger than 
the Coulomb energy $e^2/\lambda$, so that only the lowest Landau level 
enters our calculations. 
Energy spectra obtained by exact numerical diagonalization for a 
nine-electron--one-hole system are presented for $d=0$, 1.5, and 4, and 
for $\nu={1\over3}$ plus $n=0$, 1, 2, or 3 Laughlin quasielectrons (QE). 
The low-energy eigenstates can be interpreted in terms of excitonic 
complexes (or the hole) weakly interacting with the remaining electrons.

For $d\ll1$, the hole binds one or two electrons to form a neutral ($X^0$)
or charged ($X^-$) exciton. 
The $X^0$ in its ground state is effectively decoupled from the remaining 
$N-1$ electrons in a ``multiplicative'' state\cite{apalkov} whose energy 
is that of $N-1$ electrons shifted by the $X^0$ binding energy. 
In contrast, the $X^-$ is a charged Fermion, and it has Laughlin-like 
correlations with the remaining $N-2$ electrons\cite{iza,wojs1}.
For $d\gg1$, the potential of the hole is a weak perturbation on the 
eigenstates of the $N$-electron system. 
The low-lying eigenstates can be understood in terms of the angular momenta 
of the Laughlin QE's and of the hole. 
For intermediate values of $d$ ($1\leq d\leq2$), the potential of the 
hole is not strong enough to bind an electron, but it is not a weak 
perturbation on the eigenstates of the $N$-electron system, either. 
In this case the hole binds one or more Laughlin QE's to form 
fractionally charged excitons (FCX). 
We denote a bound state of the hole and $n$ QE's as $h$QE$_n$.

There are two separate symmetries which dictate the rules for radiative 
recombination of an electron--hole pair. 
The most important one is translational invariance, which in the Haldane 
spherical geometry becomes rotational invariance. 
It requires the total angular momentum $L$ and its $z$-component to be 
conserved in the radiative recombination process. 
Here $L$ denotes the total angular momentum of the system, not just 
that of the excitonic complex involved in the recombination process. 
The second symmetry is called the ``hidden symmetry''\cite{lerner}. 
It results from the fact that the commutator of the interaction 
Hamiltonian with the photoluminescence operator $\hat{\cal{L}}=
\int d^2r\,\hat{\Psi}_e(r)\hat{\Psi}_h(r)$ is proportional to 
$\hat{\cal{L}}$ whenever the magnitude of the electron--hole 
interaction $\left|V_{eh}\right|$ is equal to that of the 
electron--electron interaction $\left|V_{ee}\right|$. 
Because of this symmetry, when $d$ is equal to zero, only 
``multiplicative'' states can undergo radiative recombination. 
Therefore when $d\ll1$, the PL spectrum contains information about the 
$X^0$, while for $d>1$ it contains information about the elementary 
excitations of the Laughlin fluid and their interactions with one 
another and with the hole.

\section{Energy Spectra}

\begin{figure}
\epsfxsize=6.5in
\epsffile{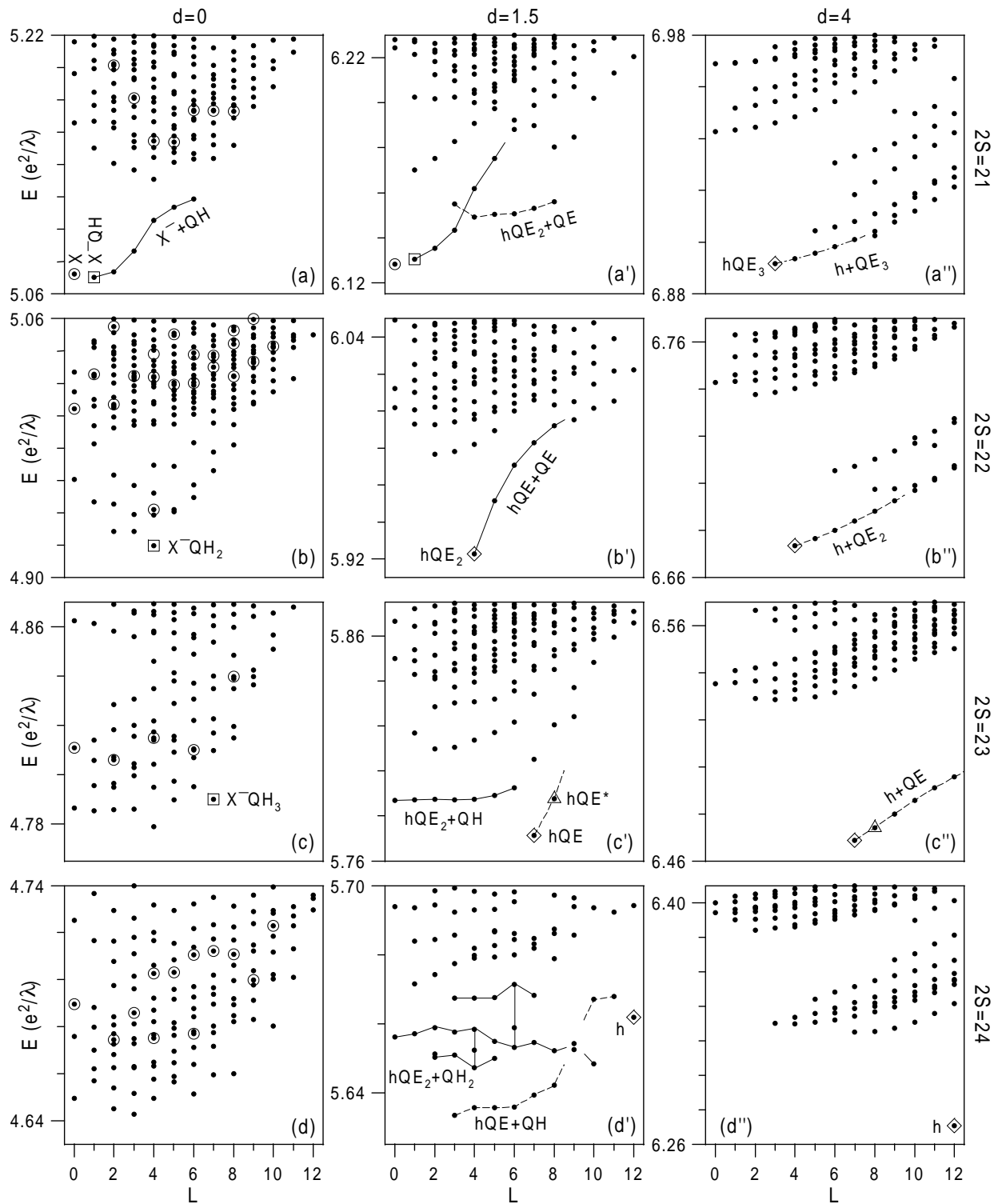}
\caption{
   Energy spectra of the nine-electron--one-hole system for the 
   monopole strength $2S=21$, 22, 23, 24 (from top to bottom), 
   and for the interplane separation $d=0$, 1.5, 4 (from left 
   to right). Lines and open symbols mark the low-energy states 
   containing different bound excitonic complexes.}
\label{fig1}
\end{figure}

In Fig.~\ref{fig1} we present the examples of energy spectra of the 
nine-electron--one-hole system obtained by exact diagonalization 
in the spherical geometry. 
The separation $d$ between the electron and hole planes is accounted 
for by taking $V_{eh}(r)=-e^2(r^2+\lambda^2d^2)^{-1/2}$. 
The radial magnetic field is given by $4\pi R^2B=2S\phi_0$, where $R$ is 
the radius of the sphere, $\phi_0=hc/e$ is the quantum of the flux, and 
the ``monopole strength'' $2S$ is equal to an integer. 
In different frames, the values of $2S$ are 21 (a--a$''$), 22 (b--b$''$), 
23 (c--c$''$), and 24 (d--d$''$), and they correspond to the Laughlin 
$\nu={1\over3}$ state with 3, 2, 1, and 0 QE excitations respectively. 
The interplanar separation equals $d=0$ (a--d), $d=1.5$ (a$'$--d$'$), 
and $d=4$ (a$''$--d$''$).

\subsection{Strong Coupling}

For $d=0$, $X^0$ and $X^-$ bound states occur. Because of the ``hidden
symmetry,'' the multiplicative states containing an $X^0$ have the 
same spectrum as the eight-electron system shifted by the $X^0$ binding 
energy. 
The CF model\cite{jain} tells us that the effective monopole strength 
seen by one CF in a system of $N'=N-1=8$ electrons near $\nu={1\over3}$ 
is $2S^*=2S-2(N'-1)$. 
$S^*$ plays the role of the angular momentum of the lowest CF shell 
(Landau level), therefore $S^*=3.5$, 4, 4.5, and 5 for the multiplicative 
states in frames (a), (b), (c), and (d) of Fig.~\ref{fig1}, respectively. 
Because the lowest shell can accommodate $2S^*+1$ CF's, it is exactly 
filled in Fig.~\ref{fig1}(a), but there are 1, 2, and 3 excess CF's for 
Fig.~\ref{fig1}(b), (c), and (d), respectively.
The excess CF's go into the next shell as Laughlin QE's with 
$l_{\rm QE}=S^*+1$, giving one QE with $l_{\rm QE}=4$ (b), two QE's each 
with $l_{\rm QE}=4.5$ (c), and three QE's each with $l_{\rm QE}=5$ (d). 
The angular momenta of the lowest band of multiplicative states are 
obtained by addition of the angular momenta of the QE excitations, 
remembering that they are identical Fermions. 
This gives $L=0$ (a), $L=4$ (b), $L=0\oplus2\oplus4\oplus6\oplus8$ (c), 
and $L=0\oplus2\oplus3\oplus4^2\oplus5\oplus6^2\oplus7\oplus8\oplus9
\oplus10\oplus12$ (d). 
These states are shown as points surrounded by a small circle in all 
frames for $d=0$. 
In the absence of QE--QE interactions (i.e.\ for mean-field CF's) all 
the states in the lowest CF band of each spectrum would be degenerate, 
but QE--QE interactions remove this degeneracy. 
Higher-energy multiplicative states that appear in the figure contain 
additional QE--QH pairs.

For the non-multiplicative states we have one $X^-$ and $N_e=N-2$ 
remaining electrons. 
The generalized CF picture\cite{iza} allows us to predict the lowest 
energy band in the spectrum in the following way. 
The effective monopole strength seen by the electrons is $2S_e^*=
2S-2(N_e-1)-2N_{X^-}$, while that seen by the $X^-$ is $2S_{X^-}^*
=2S-2N_e$. 
Here we have attached to each Fermion (electron and $X^-$) two 
fictitious flux quanta and used the mean-field approximation to describe 
the effective monopole strength seen by each particle (note that a CF 
does not see its own flux). 
The angular momentum of the lowest CF electron shell is $l_0^*=S_e^*$, 
while that of the CF $X^-$ shell is $l_{X^-}^*=S_{X^-}^*-1$\cite{wojs2}. 
For the system with $N_e=7$ and $N_{X^-}=1$ at $2S=21$, 22, 23, and 24, 
the generalized CF picture leads to: one QH with $l_{\rm QH}=3.5$ and 
one $X^-$ with $l_{X^-}$ with $l_{X^-}^*=2.5$, giving a band at $1\leq 
L\leq6$ for Fig.~\ref{fig1}(a); two QH's with $l_{\rm QH}=4$ and one 
$X^-$ with $l_{X^-}^*=3$ giving $L=0\oplus1\oplus2^3\oplus3^3\oplus4^4
\oplus5^3\oplus6^3\oplus7^2\oplus8^2\oplus9\oplus10$ for 
Fig.~\ref{fig1}(b); 
three QH's with $l_{\rm QH}=4.5$ and one $X^-$ with $l_{X^-}^*=3.5$ 
giving $L=0\oplus1^4\oplus2^6\oplus3^7\oplus\ldots$ $\oplus11^3\oplus
12^2\oplus13\oplus14$ for Fig.~\ref{fig1}(c); and four QH's with 
$l_{\rm QH}=5$ plus one $X^-$ with $l_{X^-}^*=4$ giving $L=0^3\oplus1^6
\oplus\ldots\oplus16^2\oplus17\oplus18$ for Fig.~\ref{fig1}(d). 
In the figure, we have restricted the values of $L$ and of $E$, so 
not all the states are shown.

\subsection{Weak Coupling}

For $d\gg1$, the electron--hole interaction is a weak perturbation 
on the energies obtained for the $N$-electron system\cite{chen}. 
The numerical results can be understood by adding the angular momentum 
of the hole, $l_h=S$, to the electron angular momenta obtained from the 
simple CF model. 
The predictions are: for $2S=21$ there are three QE's each with 
$l_{\rm QE}=3.5$ and the hole has $l_h=10.5$; for $2S=22$ two QE's 
each with $l_{\rm QE}=4$ and $l_h=11$; for $2S=23$ one QE with 
$l_{\rm QE}=4.5$ and $l_h=11.5$; and for $2S=24$ no QE's and $l_h=12$. 
Adding the angular momenta of the identical Fermion QE's gives $L_e$, 
the electron angular momenta of the lowest band; adding to $L_e$ the 
angular momentum $l_h$ gives the set of allowed multiplets appearing 
in the low-energy sector.
For example, in Fig.~\ref{fig1}(b$''$) the allowed values of $L_e$ 
are $1\oplus3\oplus5\oplus7$, and the multiplets at 7 and 3 have lower
energy than those at 1 and 5. 
Four low-energy bands appear at $4\leq L\leq18$, $8\leq L\leq14$, 
$6\leq L\leq16$, and $10\leq L\leq12$, resulting from $L_e=7$, 3, 5, 
and 1, respectively.

\subsection{Intermediate Coupling}

For $d\approx1$, the electron--hole interaction results in formation 
of bound states of a hole and one or more QE's. 
In the two-electron--one-hole system, the $X^0$ and $X^-$ unbind for 
$d\approx1$, but interaction with the surrounding unbound electrons 
in a larger system can lead to persistence of these excitonic states 
beyond $d=1$. 
For example, the band of states at $d=0$ in Fig.~\ref{fig1}(a) that 
we associated with an $X^-$ interacting with a QH persists at $d=1.5$ 
in Fig.~\ref{fig1}(a$'$). 
However, it appears to cross another low-energy band that extends from 
$L=3$ to 8. 
This latter band can be interpreted in terms of three QE's interacting 
with the hole as was done in the weak-coupling limit shown in 
Fig.~\ref{fig1}(a$''$). 
The other bands of the weak-coupling regime (those beginning at $L=5$, 
6, 7, 8, and 9) have disappeared into the continuum of higher states as 
a result of the increase of $V_{eh}$.

For $2S=22$, the lowest band can be interpreted in terms of one $X^-$ 
interacting with two QH's of the generalized CF picture. 
The $X^-$ has $l_{X^-}^*=3$ and the QH's each have $l_{\rm QH}=4$. 
The allowed values of $L_{2{\rm QH}}$ are 7, 5, 3, and 1, and the 
``molecular'' state QH$_2$ which has the smallest average QH--QH 
distance would have $l_{{\rm QH}_2}=7$.
This gives a band of $X^-+{\rm QH}_2$ states going from 
$L=l_{{\rm QH}_2}-l_{X^-}^*=4$ to $l_{{\rm QH}_2}+l_{X^-}^*=10$. 
A higher band might result from the 2QH state at $L_{2{\rm QH}}=5$ 
interacting with the $X^-$ to give $2\leq L\leq8$. 
The lower band beginning at $L=4$ could also be interpreted as 
a hole interacting with two QE's of the nine-electron system 
(each QE having $l_{\rm QE}=4$). 
This would produce a band of states with $4\leq L\leq18$ (arising 
from $l_{{\rm QE}_2}=7$ and $l_h=11$). 
Because the states with $L\geq8$ merge with the continuum, we 
cannot determine which of these descriptions is more appropriate 
for $d\approx1.5$ based on the energy spectra alone (to do so we 
must analyze the eigenstates).

For $2S=23$, there are two low-lying bands. 
The first contains a hole with $l_h=11.5$ and a QE with $l_{\rm QE}=4.5$. 
This gives rise to a band extending from $L=7$ to 16. 
A second band contains an additional QE--QH pair.
The cost in energy of creating this addition pair is comparable to the
energy gained through the interaction of the addition QE with the hole.
The lowest $h$QE$_2$ state occurs at $l_{h{\rm QE}_2}=l_h-l_{{\rm QE}_2}
=3.5$ (this results from choosing $l_{2{\rm QE}}=8$, the largest value 
from the set of allowed $L_{2{\rm QE}}=8$, 6, 4, 2, and 0) and adding 
$l_{h{\rm QE}_2}$ to $l_{\rm QH}=3.5$ to obtain a band with $0\leq L\leq7$. 
The state with $L=7$ in missing, undoubtedly due to the large QE--QH 
repulsion at $l_{\rm QE-QH}=1$\cite{sitko}.

For $2S=24$, the ground state at $d=1.5$ contains one hole with $l_h=12$ 
and QE--QH pair with $l_{\rm QE}=5$ and $l_{\rm QH}=4$. 
The hole and QE bind giving a set of states with $l_{h{\rm QE}}$ 
satisfying $7=l_h-l_{\rm QE}\leq l_{h{\rm QE}}\leq l_h+l_{\rm QE}=17$. 
The most strongly bound state has $l_{h{\rm QE}}=7$. 
Adding $l_{h{\rm QE}}=7$ to $l_{\rm QH}=4$ gives band $3\leq L\leq 11$ 
marked in Fig.~\ref{fig1}(d$'$). 
This band has lower energy than the Laughlin state of nine electrons and 
the hole which occurs at $L=l_h=12$.

\section{Photoluminescence}

Exact numerical diagonalization gives both the eigenvalues and the 
eigenfunctions. 
The low-energy states $\left|i\right>$ of the initial 
$N$-electron--one-hole system have just been discussed. 
The final states $\left|f\right>$ contain $N'=N-1$ electrons and no holes. 
The recombination of an electron--hole pair is proportional to the square 
of the matrix element of the photoluminescence operator $\hat{\cal{L}}$. 
We have evaluated $\left|\left<f|\hat{\cal{L}}|i\right>\right|^2$ for 
all of the low-lying initial states and have found the following results
\cite{wojs3}. 
(i) Conservation of the total angular momentum $L$ is at most weakly 
violated through the scattering of ``spectator'' particles (electrons 
or quasiparticles) which do not participate directly in the recombination 
process if the filling factor $\nu$ is less than or equal to approximately 
${1\over3}$. 
(ii) In the strong-coupling region, the neutral $X^0$ line is the dominant 
feature of the PL spectrum. 
The $X^-{\rm QH}_2$ state has very small oscillator strength for 
radiative recombination. 
(iii) For intermediate coupling, the $h$QE$_2$ and an excited state of 
the $h$QE (which we denote by $h$QE$^*$) are the only states with large 
oscillator strength for photoluminescence.

At zero temperature ($T=0$), all initial states must be ground states 
of the $N$-electron--one-hole system. 
At finite but low temperatures, excited initial states contribute to 
the PL spectrum. 
The photoluminescence intensity is proportional to
\begin{equation} 
   w_{i\rightarrow f}=
   \frac{2\pi}{\hbar}{\cal Z}^{-1}
   \sum_{i,f}e^{-i\beta E_i}
   \left|\left<f\right|\hat{\cal{L}}\left|i\right>\right|^2 
   \delta(E_i-E_f-\hbar\omega),
\nonumber
\end{equation}
where $\beta^{-1}=k_BT$ and ${\cal Z}=\sum_ie^{-\beta E_i}$.
It is worth illustrating how the $h$QE$_2\rightarrow{\rm QH}+\hbar\omega$
process satisfies the $\Delta L=0$ selection rule. 
An initial state containing one hole and two QE's of an $N$-electron 
system must have $2S=3(N-1)-2=3N-5$.
Each QE will have $l_{\rm QE}={1\over2}(N-1)$ and the hole has 
$l_h=S=\frac{3}{2}N-\frac{5}{2}$. 
The most strongly bound $h$QE$_2$ state has $l_{{\rm QE}_2}=2l_{\rm QE}-1
=N-2$ and $l_{h{\rm QE}_2}=l_h-l_{{\rm QE}_2}=\frac{1}{2}(N-1)$.
The final state contains $N'=N-1$ electrons and a single QE with
$l_{\rm QH}=S-(N'-1)=\frac{1}{2}(N-1)$. 
Thus the initial and final states each have $L=\frac{1}{2}(N-1)$, 
so the $\Delta L=0$ selection rule is satisfied. 
The same thing can be done for the excited $h{\rm QE}^*$ complex
(with $L$ larger by one unit than the ground state of the $h{\rm QE}$). 
It satisfies the $\Delta L=0$ selection rule, but the ground state 
$h{\rm QE}$ does not.

The authors would like to acknowledge the support of the Materials 
Research Program of Basic Energy Sciences, US Department of Energy.


\end{document}